\documentclass[conference]{IEEEtran}
\IEEEoverridecommandlockouts
\usepackage[T1]{fontenc}
\usepackage{graphicx}
\usepackage{cite}
\usepackage{bm}
\usepackage{url}
\usepackage{amsmath}
\usepackage{amssymb}
\usepackage{xcolor}
\usepackage{amsfonts}
\usepackage{amsthm}
\usepackage{algorithm2e}

\newtheorem{theorem}{Theorem}

\def\BibTeX{{\rm B\kern-.05em{\sc i\kern-.025em b}\kern-.08em
    T\kern-.1667em\lower.7ex\hbox{E}\kern-.125emX}}
\begin{document}

\title{Physics-Aware Linearized ADMM and Its Unrolling
\thanks{This work is partly supported by JST, CRONOS, Japan Grant Number JPMJCS25N5.}
}

\author{\IEEEauthorblockN{Satoshi Takabe\IEEEauthorrefmark{1}\IEEEauthorrefmark{2}}
\IEEEauthorblockA{\IEEEauthorrefmark{1}\textit{Department of Computer Science } \\
\textit{Institute of Science Tokyo}\\
Tokyo, Japan \\
takabe@comp.isct.ac.jp, arai.s.ba03@m.isct.ac.jp }\and
\IEEEauthorblockN{Shunta Arai\IEEEauthorrefmark{1}}
\IEEEauthorblockA{\IEEEauthorrefmark{2}\textit{Center for Advanced Intelligence Project} \\
\textit{RIKEN}\\
Tokyo, Japan}\and\IEEEauthorblockN{Tadashi Wadayama\IEEEauthorrefmark{3}}
\IEEEauthorblockA{\IEEEauthorrefmark{3}\textit{Department of Engineering} \\
\textit{Nagoya Institute of Technology}\\
Nagoya, Japan \\
wadayama@nitech.ac.jp}
}

\maketitle

\begin{abstract}
Recently, partial differential equations (PDEs) have been used to directly model the measurement process in signal processing, although their  evaluation is costly.
In this paper, we propose a novel alternating direction method of multipliers (ADMM)-based algorithm called physics-aware linearized ADMM (PA-LADMM) for inverse problems from PDE-based measurement processes.
The key idea is the linearization of the subproblem with PDEs, leading to a cost-efficient update rule that calls only a PDE solver and its gradient evaluation per iteration. The algorithm has a theoretical convergence guarantee under certain conditions.
In addition, we combine it with deep unfolding to unroll the PA-LADMM and train its internal parameters using supervised data.
Two distinct experiments, compressed sensing with optical fiber communication and image restoration from noisy anisotropic diffusion, demonstrated the effectiveness of the proposed algorithms.
\end{abstract}


\section{Introduction}\label{sec:intro}

In signal processing, inverse problems are fundamental, where the goal is to recover an unknown signal from a noisy measurement. 
It has been common to formulate the inverse problem based on the prior information of the original signal, such as  
compressed sensing (CS)~\cite{CS} for sparse signal recovery.
On the other hand, modeling of a measurement process usually depends on its statistical properties for simplicity and tractability, whereas such measurements are often based on the first principles of physics governed by partial differential equations (PDEs).
As an example, the propagation of electromagnetic waves is governed by Maxwell's equations.

Recently, emulating physical systems or solving PDEs has been of interest in the field of deep learning. 
For example, physics-informed neural networks (PINNs)~\cite{PINN} have been proposed to solve PDEs by minimizing the residual of the PDE and initial and boundary conditions. They can predict the solution of PDEs accurately without solving them in advance. 
Related to signal processing,~\cite{Song2024} proposed the alternating direction method of multipliers (ADMM)-based PINN to solve the inverse problem whose measurement process is defined by a PDE.
This direction is promising because the PINN can predict a solution of the PDE in a mesh-free manner. 
However, it requires training separate PINNs for each iteration, which can be computationally demanding.

Another direction is the use of a differentiable PDE solver~\cite{DiffPDE} which usually uses a finite difference method. Those solvers enable us to use auto-differentiation, which is useful for, e.g., trainable parameters in NNs via backpropagation~\cite{Kag2022}.
In this direction, Wadayama \textit{et al.}~\cite{WIT} proposed the physics-aware iterative shrinkage-thresholding algorithm (PA-ISTA) using deep unfolding (DU)~\cite{LISTA}. "Physics-aware" emphasizes the use of a PDE solver in the algorithm, as opposed to PINNs.
DU (or algorithm unrolling) is a deep-learning technique to tune an iterative algorithm by unrolling its structure and embedding trainable parameters. 
DU has been successfully applied to various algorithms for inverse problems such as CS~\cite{TISTA} and image restoration~\cite{ISTANet}.
The advantage of DU-based algorithms is the low training cost thanks to the small number of trainable parameters, which is preferable for practical system models with complex PDEs.
However, the previous study~\cite{WIT} was limited to the CS problem and ISTA. 

In this paper, we extend the PA framework to ADMM. ADMM is a natural choice for several reasons~\cite{Boyd}: (i) it often exhibits faster convergence than gradient-based methods like ISTA; (ii) it separates the data fidelity term and regularization term, enabling flexible regularizer selection. 
However, directly applying ADMM to PDE-based inverse problems is computationally expensive because it requires many gradient evaluations through a PDE solver.

To address this, we propose a novel ADMM-based algorithm called PA-linearized ADMM (PA-LADMM) inspired by a linearization technique for ADMM~\cite{LADMM}.
The proposed algorithm is a general signal recovery algorithm for inverse problems with a PDE-based measurement process and a wide class of regularizers, and has a provable convergence guarantee.
In addition, we propose a DU-PA-LADMM whose internal parameters are learned using the DU framework.
The performance of the proposed algorithms is demonstrated by numerical experiments on two distinct problems: CS with optical fiber communication and image restoration from noisy anisotropic diffusion.

In the rest of this paper, Sec.~\ref{sec:problem} formulates the problem. Section~\ref{sec:ADMM} introduces the proposed PA-LADMM and Sec.~\ref{sec:DU-LADMM} proposes its training based on DU. Section~\ref{sec:exp} demonstrates numerical experiments. Section~\ref{sec:conclusion} concludes the paper.

\textit{Notation}: We use $[n]:=\{1,2,\dots,n\}$. 
A differentiable function $f(\bm x):\mathbb{R}^d \to \mathbb{R}$ is said to be $L$-differentiable if its derivative is $L$-Lipschitz continuous, i.e., $\|\nabla f(\bm x) - \nabla f(\bm x')\|_2 \leq L \|\bm x - \bm x'\|_2$ for all $\bm x, \bm x' \in \mathbb{R}^d$.
For a convex function $r(z):\mathbb{R}^d \to \mathbb{R}$, the proximal operator is defined by 
$\mathrm{prox}_{r}(\bm x) := \arg\min_{\bm z} \{r(\bm z) + \|\bm z - \bm x\|_2^2/2\}$. 
$\mathcal N(\bm \mu,\bm \Sigma)$ denotes the multivariate Gaussian distribution with mean $\bm \mu$ and covariance matrix $\bm \Sigma$.

\section{Problem Setting} \label{sec:problem}

In this paper, we consider the following (possibly nonlinear) measurement:
$
\bm y = F(\bm x) + \bm w
$,
where $\bm x \in \mathbb{R}^d$ is an original signal, 
$\bm w \sim \mathcal{N}(\bm 0, \sigma^2 \bm I)
$ is a random noise vector with parameter $\sigma>0$, and $\bm y \in \mathbb{R}^m$ is the corresponding measured signal.
In addition, $F(\bm x):\mathbb{R}^d \to \mathbb{R}^m$ is an observation defined by a given PDE with initial condition $\bm x$. 
We assume that $F$ has no analytic form and its numerical solution is obtained by a differentiable PDE solver.

The goal is to recover the original signal $\bm x$ from the measured signal $\bm y$.
We thus consider the following regularized least squares problem:
\begin{align}
\min_{\bm x} \frac{1}{2} \|F(\bm x) - \bm y\|_2^2 + R(\bm x),
\label{eq:reg}
\end{align}
where $R(\cdot)$ is a regularizer for the prior knowledge on $\bm x$.

\section{ADMM and its linearization}\label{sec:ADMM}
In this section, we first formulate the ADMM for the problem \eqref{eq:reg} and then linearize it to reduce the computational cost for solving the subproblem.

To derive the ADMM update rules, \eqref{eq:reg} is rewritten as:
\begin{align}
\min_{\bm x, \bm z} \frac{1}{2} \|F(\bm x) - \bm y\|_2^2 + R(\bm z) \mbox{ s.t. } \bm x = \bm z,
\label{eq:reg2}
\end{align}
by introducing the auxiliary variable $\bm z \in \mathbb{R}^d$.

Then, we define the augmented Lagrangian function:
\begin{align}
\mathcal{L}(\bm x, \bm z, \bm u) &\!=\! \frac{1}{2} \|F(\bm x) \!-\! \bm y\|_2^2 \!+\! R(\bm z) 
 \!+\! \frac{\rho}{2} \|\bm x \!-\! \bm z\|_2^2 \!+\! \bm u^T (\bm x \!-\! \bm z),
\nonumber 
\end{align}
where $\bm u \in \mathbb{R}^d$ is a dual variable and $\rho>0$ is a penalty parameter. 
The ADMM update rules are given by:
\begin{align}
\bm z_{k+1} &= \arg\min_{\bm z} \mathcal{L}(\bm x_{k}, \bm z, \bm u_k), \label{eq:ADMM_z}\\
\bm x_{k+1} &= \arg\min_{\bm x} \mathcal{L}(\bm x, \bm z_{k+1}, \bm u_k), \label{eq:ADMM_x}\\
\bm u_{k+1} &= \bm u_k + \rho (\bm x_{k+1} - \bm z_{k+1}). \label{eq:ADMM_u}
\end{align}
The drawback of this standard ADMM is that the subproblem \eqref{eq:ADMM_x} has no analytic form in general because of $F(\bm x)$. 
For instance, a gradient method for the subproblem requires the evaluation of the gradient of $F(\bm x)$ through the solver iteratively, which is computationally costly.

To address this issue, we use the LADMM by linearizing $F(\bm x)$ around $\bm x_{k}$~\cite{LADMM}.
Then, the cost function of \eqref{eq:ADMM_x} is replaced by the following:
\begin{align}
(\bm x\!-\!\bm x_k)^T\nabla F(\bm x_k) \!+\! \frac{\rho}{2} \|\bm x\!-\!\bm z_{k+1}\|_2^2  \!+\! \bm u_k^T \bm x \!+\! \frac{L_x}{2} \|\bm x\!-\!\bm x_{k}\|_2^2, \nonumber
\end{align}
where $L_x>0$ is a penalty parameter of the regularization for $\bm x$.
To ensure the convergence of the algorithm, we also introduce a regularization term with respect to $\bm z$. For $L_z>\rho$, the cost function of \eqref{eq:ADMM_z} is replaced by
\begin{align}
    R(\bm z) + \frac{\rho}{2} \|\bm x - \bm z\|_2^2 + \bm u^T (\bm x - \bm z) + \frac{L_z-\rho}{2}\|\bm z-\bm z_k\|_2^2. \nonumber
\end{align}
    
Since these have closed-form solutions, the LADMM update rules~\footnote{The order of the update rules is based on the original LADMM~\cite{LADMM}.} for PDE-based observation are therefore given by:
\begin{align}
\bm z_{k+1} &= \mathrm{prox}_{{R}/{L_z}}\left[\left(1-\frac{\rho}{L_z}\right)\bm z_k + \frac{\rho}{L_z}\bm x_{k} + \frac{\bm u_k}{L_z}\right], \label{eq:LADMM_z}\\
\bm x_{k+1} &=\frac{1}{L_x+\rho} \left(-\nabla F(\bm x_k)+L_x\bm x_k + \rho\bm z_{k+1} - \bm u_k \right), \label{eq:LADMM_x}\\
\bm u_{k+1} &= \bm u_k + \rho (\bm x_{k+1} - \bm z_{k+1}). \label{eq:LADMM_u}
\end{align}
We call it the PA-LADMM, which is a special case of the original LADMM in \cite{LADMM}.

The PA-LADMM has a provable convergence guarantee.
\begin{theorem}\label{thm:LADMM}
Assume that the data fidelity term $\|\bm y-F(\bm x)\|_2^2$ is $L_s$-differentiable,  
$R(\bm x)$ is bounded below, and the cost function of \eqref{eq:reg} diverges to $+\infty$ for any $\|\bm x\|_2\to\infty$.
Then, if the parameters of the LADMM satisfy $L_x\ge L_s+L_s^2+3$, 
$L_z\ge \rho + L_s+6L_s^2+1$, 
$\rho\ge \max\{L_s+L_x+2,6(L_s^2+L_x^2)/(L_s^2+L_x),3L_x^2\}$,
the sequence $\{L(\bm x_k, \bm z_k, \bm u_k)\}$ updated by \eqref{eq:LADMM_z}--\eqref{eq:LADMM_u} is convergent, and $\|\bm x_{k+1}-\bm x_k\|_2\to 0$, $\|\bm z_{k+1}-\bm z_k\|_2\to 0$, $\|\bm u_{k+1}-\bm u_k\|_2\to 0$, as $k\to\infty$.
\end{theorem}
We omit the proof of the theorem here because it directly follows from that of Thm.~1 in \cite{LADMM}~\footnote{The correspondence between LADMM~\cite{LADMM} and the PA-LADMM is as follows: $(\bm x,\bm y,\bm\gamma, \beta)\leftarrow (\bm z,\bm x,\bm u,\rho)$, $(\bm A,\bm B)\leftarrow(-\bm I,\bm I)$, and $(f,g,h)\leftarrow(R,\|F(\bm x) - \bm y\|_2^2,0)$.}.

In addition, the fixed point of the LADMM is a saddle point of the problem \eqref{eq:reg}.
However, the convergence performance and fixed point depend on the parameters $L_x, L_z, \rho$ of the LADMM, whose tuning is a practical issue in signal recovery.

\section{DU-PA-LADMM}\label{sec:DU-LADMM}
In this section, we apply the DU scheme to the PA-LADMM to learn the internal parameters $L_x, L_z, \rho$ using supervised data.
A standard DU scheme unrolls the iterative structure of the differentiable algorithm like gradient descent.
Then, by embedding trainable parameters into the unrolled network, we can train the parameters using a standard supervised learning framework such as backpropagation.
In the case of the LADMM, the trainable parameters $\{L_x^{(k)}, L_z^{(k)}, \rho^{(k)}\}$ 
 are embedded in its update rules in \eqref{eq:LADMM_z}--\eqref{eq:LADMM_u} for $k=0,\dots,K-1$.

 \RestyleAlgo{ruled}
 \begin{algorithm}[t]
 \caption{Learning protocol of PA-LADMM }\label{alg:1}
  \SetKwInOut{Input}{Input}
 \SetKwInOut{Output}{Output}
 \SetKwInOut{Learning}{Learning}
 \SetKwInOut{Test}{Test}
 \SetKwFor{For}{for}{ }{}
 \SetKwInOut{Initialize}{Initialize}
  \Input{$K$ (number of iterations), $N_{\mathrm{epoch}}$ (number of epochs), $\mathcal D$ (training data)}
 \For{$e = 0,\dots,N_{\mathrm{epoch}}-1$}{
     \For {each mini-batch in $\mathcal D$}{
     \text{$\mathcal{F} \leftarrow \emptyset$}\\
     \For{$k = 0,\dots,K-1$ \hfill{// Store phase}}{
         \text{Calculate $\|F(\bm x_k)-\bm y\|_2^2$ by the PDE solver.}\\
         \text{Calculate its gradient by backpropagation.}\\
         \text{and add it to $\mathcal{F}$.}\\
         \text{Update $(\bm x_{k+1}, \bm z_{k+1}, \bm u_{k+1})$ by 
         \eqref{eq:LADMM_z}--\eqref{eq:LADMM_u}.}
         }     
     \For {$k = 0,\dots,K-1$\hfill{// Replay phase}}{
         \text{Update $(\bm x_{k+1}, \bm z_{k+1}, \bm u_{k+1})$ by \eqref{eq:LADMM_z}--\eqref{eq:LADMM_u}}\\
         \text{using gradient stored in $\mathcal{F}$.}
     }
     \text{Update trainable parameters by backpropagation}
     }
 }
  
 \end{algorithm}

However, unlike the usual recipe, we need to calculate the gradient of $F$ using backpropagation through a PDE solver in the forward pass. 
It prevents the direct application of backpropagation in the training phase of the algorithm.
To mitigate this issue, it has been proposed to execute the algorithm twice in the forward pass~\cite{WIT}: once for storing the gradient information through the PDE solver, and once for the execution of the algorithm and backpropagation in the backward pass.
The pseudocode of the proposed DU-PA-LADMM is given in Alg.~\ref{alg:1}.
The key idea is to pre-compute and store the gradient information of $F$ in ``store phase''. It enables us to execute backpropagation in the backward pass and update the trainable parameters in the ``replay phase.''

\section{Experiments}\label{sec:exp}
In this section, we evaluate the performance of the proposed (DU-)PA-LADMM on the two distinct problems: CS with optical fiber communication and image restoration with anisotropic diffusion.
We focus on the performance of the proposals with a finite number of iterations. The tuned or trained parameters may violate the conditions of Thm.~\ref{thm:LADMM} but have empirically achieved better results.

\subsection{Compressed sensing with optical fiber communication}

The setting is based on the previous work in~\cite{WIT}.
The optical field in the single-mode optical fiber is described by the nonlinear Schrödinger equation~\cite{SSFM} defined by
\begin{align}
\frac{\partial U}{\partial s} =  \frac{\mathrm{i}\beta_2}{2}\frac{\partial^2 U}{\partial t^2} + \mathrm{i}\gamma|U|^2 U,
\label{eq:nlse}
\end{align}
where $U=U(s,t)\in \mathbb{C}$ is the optical field, $s$ is the position along the fiber, and $t$ is the time.
The second term on the right-hand side represents the nonlinearity of the PDE.

Here, we consider the sparse signal recovery in the optical fiber communication.
We assume that the input at $s=0$ is a superposition of Gaussian pulses, which is given by
\begin{align}
U(0,t) = \sum_{i=1}^n x_i \exp\left(-\frac{(t-t_i)^2}{2\sigma_t^2}\right),
\label{eq:initial}
\end{align}
where $\bm x =(x_1,x_2,\ldots,x_n)^T \in \mathbb{C}^n$ is a sparse signal, $t_i$ ($i\in[n]$) is an input time, and $\sigma_t$ is the standard deviation of a pulse.
Then, we assume that the detector directly observes the output waveform $y_j = U(q_j;\bm x) + w_j$ at $s=L$ at time $t=q_j$ ($j\in[m]$), 
where $U(q_j;\bm x)$ is $U(L,q_j)$ with the initial condition $\bm x$ and $w_j$ is a complex Gaussian noise.

The goal is to recover the sparse signal $\bm x$ from the measured signal $\bm y$, which is directly formulated by \eqref{eq:reg} with $\ell_1$ norm $R(\bm x) = \| \bm x\|_1$.
Since $\bm x\in \mathbb{C}^n$, we need to extend the problem and the LADMM to the complex domain by regarding inner products as Hermitian inner products.
Then, the proximal operator in \eqref{eq:LADMM_z} is the (complex) soft-thresholding function with a parameter $\tau$ given by
$T_{\tau}(x) := x/|x|\max\{|x|-\tau,0\} $.
In addition, the derivative in the PA-LADMM is regarded as the Wirtinger derivative.
For a function $f(x):\mathbb{C} \to \mathbb{R}$, a Wirtinger derivative~\cite{Wirtinger} is defined by
\begin{equation}
\frac{\partial f}{\partial x^\ast} := \frac{1}{2} \left(\frac{\partial f}{\partial \Re(x)} + \mathrm{i}\frac{\partial f}{\partial \Im(x)}\right).\label{eq:wirtinger}
\end{equation}
It is known that the steepest descent direction of $f(x)$ is given by ${\partial f}/{\partial x^\ast}$~\cite{CompGrad}. We thus replace the gradient $\nabla F(\bm x_k)$ in the PA-LADMM update rule~\eqref{eq:LADMM_x} by \eqref{eq:wirtinger}.

In the experiment, we assumed a typical propagation condition~\cite{WIT}. 
We set $\beta_2=-10$, $\gamma=2$ for \eqref{eq:nlse}.
The time evolution of \eqref{eq:nlse} is numerically solved by the split-step Fourier method (SSFM)~\cite{SSFM} with the initial condition \eqref{eq:initial}.
The step sizes of SSFM were set to $\Delta s = 0.01$ for $s$-direction and $\Delta t=0.3$ for the time interval $t\in [-38.4, 38.4]$.
The parameters for sparse signal~\eqref{eq:initial} were set to $t_i=2i-31$ ($i\in[n]$) with $n=30$ and $\sigma_t=1$, resulting in a dispersion length $L_d:=\sigma_t^2/|\beta_2|=0.1$.
In addition, we assumed that randomly chosen $3$ elements take uniformly distributed random complex values with unit norm in $\bm x$ of length $n=30$. The remaining elements were set to zero.       
For the observation, we set $L=3L_d=0.3$ and  $q_j=-38.4+j\Delta t$ for $j\in[m]$, where $m=256$. 
The signal-to-noise ratio (SNR) was set to $15$~dB.

The parameters of the PA-LADMM were manually tuned to $(L_x,L_z,\rho)=(125,7,2)$.
In the training of the DU-PA-LADMM, we employed mini-batch training with a batch size of $32$ and $300$ epochs. The Adam optimizer~\cite{Adam} with a learning rate of $2\times 10^{-4}$ was used for minimizing the loss function.
We set $K=30$ and used the weighted mean squared error (MSE) loss function given by
\begin{align}
    \mathcal U := \frac{1}{2}\left\| \bm y - \bm U(\bm x_K) \right\|_2^2+  \frac{\alpha}{2} \sum_{k=1}^{K-1} \left\| {\bm y - \bm U(\bm x_k)} \right\|_2^2, \nonumber
\end{align}
where $\bm x_k$ is the estimate at the $k$-th iteration 
and $\bm U(\bm x):=(U(q_1;\bm x),\dots, U(q_m;\bm x))^T$.
The weight $\alpha$ was set to $0.01$. This MSE loss evaluated the final estimate and intermediate signals to make the training stable~\cite{Guide}.
The initial values of trainable parameters $\{L_x^{(k)}\}$, $\{L_z^{(k)}\}$, and $\{\rho^{(k)}\}$ were set to $10$, $100$, and $1$, respectively. 
All trainable parameters were rescaled; e.g., $\tilde L_k=\log_{10}{L_k}$ because their values span a wide range.
The initial $\bm x_0$ was set to an estimate by the conventional digital back propagation (DBP)~\cite{DBP}, which worked well in the noiseless case.
The code was written in Julia.

For baselines, we used the DBP and the PA-ISTA with $30$ iterations, whose parameters were trained under the same conditions as in~\cite{WIT}. 

\begin{figure}[t]
    \centering
    \includegraphics[width=0.4\textwidth]{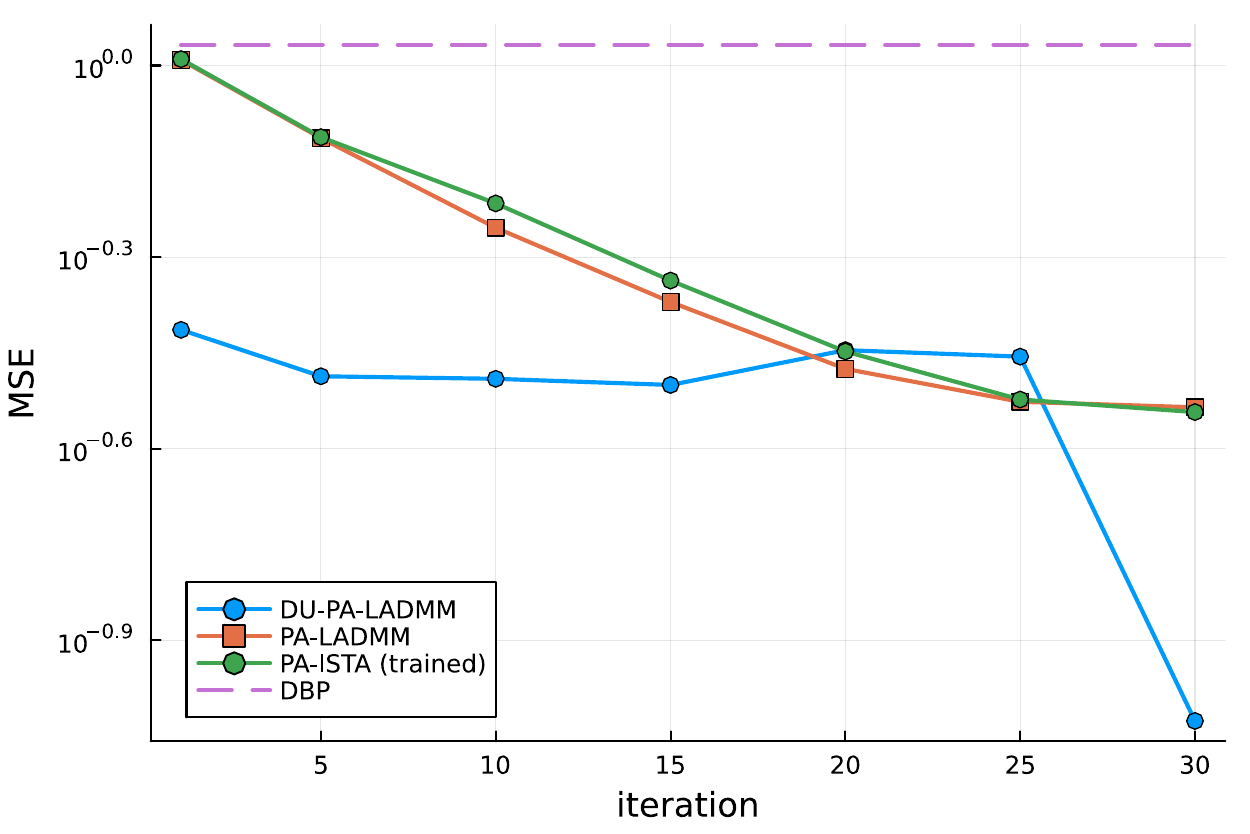}
    \caption{MSE performance as a function of iterations when $(n,m)=(30,256)$ and SNR$=15$~dB. MSE was averaged over $100$ random samples.}
    \label{fig:opt_MSE}
\end{figure}

Figure~\ref{fig:opt_MSE} shows the comparison of the MSE of the reconstructed signals by the proposed (DU-)PA-LADMM and the baselines.
We found that the DU-PA-LADMM achieved the best performance, whereas the PA-LADMM showed similar performance to the PA-ISTA. The performance improvement over the PA-LADMM shows the effectiveness of the DU scheme.
Note that the non-monotonic behavior of MSE of the DU-PA-LADMM is often seen for DU-based methods~\cite{Guide}.

\begin{figure}[t]
    \centering
    \includegraphics[width=0.4\textwidth]{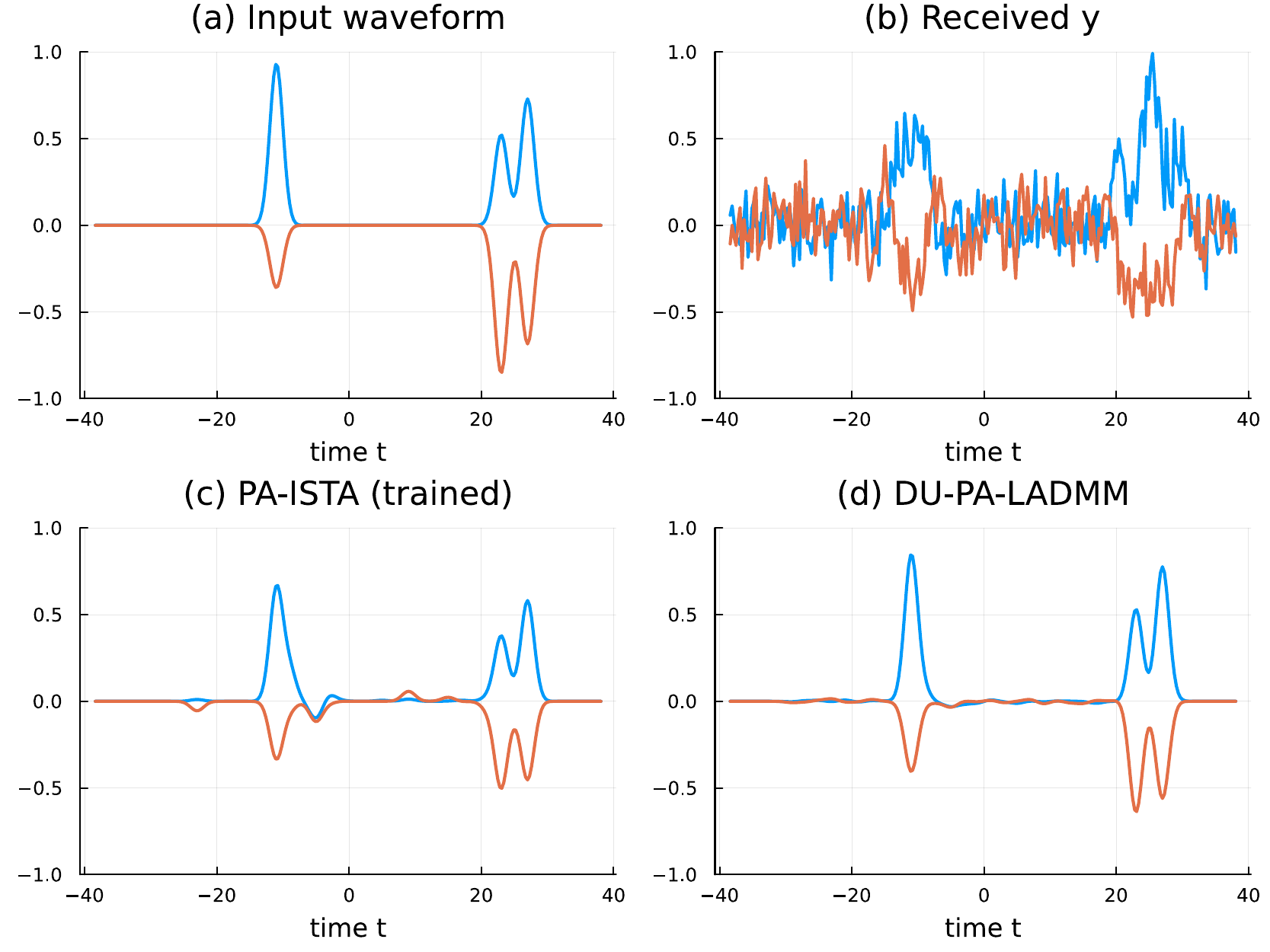}
    \caption{Comparison of waveforms (blue: real part, red: imaginary part). (a) Input waveform as the ground truth. (b) Observed waveform. Reconstructed waveforms by (c) trained PA-ISTA and (d) DU-PA-LADMM.}
    \label{fig:opt_wave}
\end{figure}

Figure~\ref{fig:opt_wave} shows the comparison of the reconstructed waveforms by the proposed DU-PA-LADMM and PA-ISTA.
Although both methods recovered sparse signals, the PA-ISTA exhibited small spurious peaks, whereas the DU-PA-LADMM produced a cleaner waveform.
In addition, the DU-PA-LADMM estimated the intensity of non-zero $x_i$'s to be closer to the input, which was not the case for the PA-ISTA.

\subsection{Image restoration from noisy anisotropic diffusion}
In this subsection, we consider another example of PDE-based observation. 
Anisotropic diffusion is an image processing technique that is used to smooth images while preserving edges.
Here, we regard it as a nonlinear blur and attempt to recover the original image from a noisy blurred image.

For each color channel of an image, anisotropic diffusion is formulated by the Perona-Malik (PM) equation~\cite{PM}:
\begin{align}
    \frac{\partial I}{\partial t} = \nabla \cdot (c(\nabla I) \nabla I), \label{eq:PM}
\end{align}
where $I=I(a,b,t)$ is the normalized intensity at location $(a,b)$ and $t$ is the time, and 
$c(\bm v) = \exp(-\|\bm v\|_2^2/\kappa^2)$ is a diffusion coefficient with $\kappa >0$.
Then, the observation is defined by 
\begin{align}
    y_{a,b} = I(a,b,T; \bm x) + w_{a,b}, \quad (a\in[H], b\in[W]) \nonumber
\end{align}
where $I(a,b,T; \bm x)$ is the solution of \eqref{eq:PM} with the initial condition $I(a,b,0) = x_{a,b}$ and $w_{a,b} \sim \mathcal{N}(0, \sigma^2)$ is a noise.

To recover the original image, we consider the optimization problem inspired by the Wavelet thresholding (WT) algorithm for image denoising~\cite{WT}.
The problem is given by
\begin{align}\label{eq:PM_opt}
    \min_{\bm x} \frac{1}{2} \|\bm I(\bm x,T) - \bm y\|_2^2 + R( \psi(\bm x)),
\end{align}
where $(\bm I(\bm x,T))_{a,b} = I(a,b,T; \bm x)$ and $\psi(\cdot)$ is the Wavelet transform.
Note that, if $\bm I = \bm x$, the solution corresponds to the original WT algorithm~\cite{WT_prox}, which is given by 
    $\bm {\hat x} ^{\mathrm{WT}} = \psi^{-1}(\eta_{\tau}( \psi(\bm x) ))$, 
where $\psi^{-1}(\cdot)$ is the inverse Wavelet transform and $\eta_{\tau}(x) := x/|x|\max\{|x|-\tau,0\}$ is the (real) soft-thresholding function with a parameter $\tau$.
In the case of PA-LADMM, the update rule~\eqref{eq:LADMM_z} is given by
\begin{equation*}
\bm z_{k+1} = \psi^{-1}\left(\eta_{1/{L_z}}\left[\psi\left(\left(1-\frac{\rho}{L_z}\right)\bm z_k + \frac{\rho}{L_z}\bm x_{k} + \frac{\bm u_k}{L_z}\right)\right]\right).
\end{equation*}

In the experiment, we attempted the recovery of colored images of size $H\!=\!W\!=\!32$. The code was written in PyTorch.
In CIFAR-10 dataset~\cite{CIFAR10}, $960$ images were randomly selected for training and $160$ images for testing.
As an image degradation model, we used the PM equation~\eqref{eq:PM} with $\kappa=T=1.0$ and SNR of $25$~dB. 
In implementation, the equation was solved by the finite difference method with time step $\Delta t = 0.05$. 
We used the $2$-level Haar Wavelet transform as $\psi(\cdot)$.
For the training of the DU-PA-LADMM, we employed mini-batch training with a batch size of $32$ and $50$ epochs with early stopping. The Adam optimizer~\cite{Adam} with a learning rate of $1\times 10^{-3}$ was used for minimizing the MSE loss function between the output of the DU-PA-LADMM and the original image. 
The number of iterations was set to $K=15$.
The initial values of trainable parameters $\{L_x^{(k)}\}$, $\{L_z^{(k)}\}$, and $\{\rho^{(k)}\}$ were set to $5$, $25$, and $0.01$, respectively.
All trainable parameters were rescaled; e.g., $\tilde L_k=\log_{10}{L_k}$.

For baselines, we used Gaussian smoothing~\cite{Gauss} ("Gauss") with $\sigma=0.5$ and the bilateral filter~\cite{Bilateral} ("Bilateral") with $\sigma_s=1.0$ and $\sigma_r=0.02$ as conventional filters.
For ablation studies, we used WT with $\tau =0.01$, which neglects the anisotropic diffusion. We also used the gradient descent (GD) for \eqref{eq:PM_opt} with $R(\cdot)\equiv 0$, which neglects the regularization term.
The step size of GD was set to $0.1$ and the number of iterations was set to $15$. These parameters were manually tuned.

Table~\ref{tab:results} shows the quantitative comparison of the image restoration methods. 
As performance metrics, we used peak SNR ("PSNR"), structural similarity index ("SSIM")~\cite{SSIM}, and learned perceptual image patch similarity ("LPIPS")~\cite{LPIPS}, which are commonly used for performance evaluation. 
We found that conventional methods such as Gaussian smoothing, the bilateral filter, and WT showed poor performance. In particular, these methods increased LPIPS, implying restored images perceptually differed from the original image due to neglecting the PDE-based measurement.
On the other hand, PA methods achieved better performance in all metrics. In particular, the proposals were superior to GD, indicating the WT-based denoising was effective. The trained DU-PA-LADMM achieved the best performance among all methods, demonstrating the effectiveness of the combination of the PA-LADMM and DU.

\begin{figure}[t]
    \centering
    \includegraphics[width=0.45\textwidth]{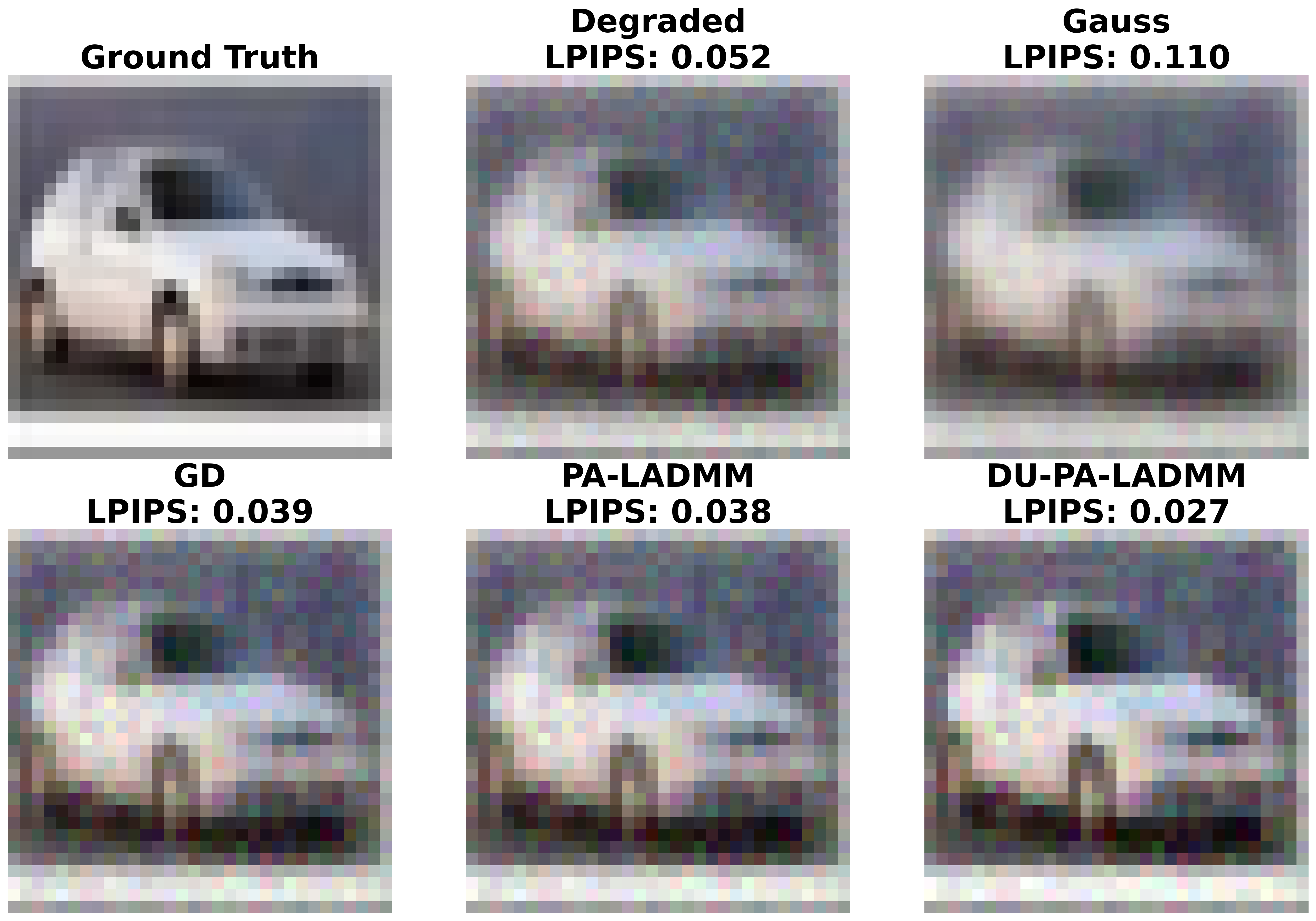}
    \caption{Comparison of restored images with LPIPS scores.}
    \label{fig:PM_comp}
\end{figure}

Figure~\ref{fig:PM_comp} shows an example of a restored image with an LPIPS score. We found that the DU-PA-LADMM successfully removed the blur of the image while keeping its edges sharp.


\begin{table}[t]
\centering
\caption{Performance comparison of image restoration methods for anisotropic diffusion.}
\label{tab:results}
\begin{tabular}{lccc}
\hline
Method & PSNR [dB] ($\uparrow$) & SSIM ($\uparrow$) & LPIPS ($\downarrow$) \\
\hline
Degraded & 24.17  & 0.788 & 0.0587 \\
Gauss & 24.02 & 0.788 & 0.0943 \\
Bilateral & 23.70 & 0.762 & 0.0640 \\
WT & 24.15 & 0.789 & 0.0677 \\
GD & 25.05 & 0.815 & 0.0402 \\
PA-LADMM & 25.23 & 0.820 & 0.0388 \\
DU-PA-LADMM & \textbf{25.77} & \textbf{0.839} & \textbf{0.0279} \\
\hline
\end{tabular}
\end{table}

\section{Conclusion}\label{sec:conclusion}
In this paper, we proposed the LADMM algorithm for signal recovery from a PDE-based measurement process.
By utilizing linearization of the subproblem, the proposed algorithm consists of cost-efficient update rules. The convergence to a saddle point of the algorithm is guaranteed, which is an advantage of the LADMM. 
In addition, by combining it with DU, we proposed a trainable PA-LADMM algorithm called DU-PA-LADMM.
Two distinct experiments, CS with optical fiber communication and image restoration from noisy anisotropic diffusion, showed the effectiveness of the proposals. The PA-LADMM showed better performance than GD-based methods, whereas DU further improved the performance by training its internal parameters.

There are several directions for future work. 
The application of the proposed algorithm to other settings in signal processing is promising to enhance the effectiveness of PA methods in signal processing. 
Another is the extension of the proposals to distributed optimization, which will be useful for sensing systems with multiple sensors.

\end{document}